\documentclass[12pt]{article}
\usepackage{graphics} 
\def\lsim{\mathrel{\lower2.5pt\vbox{\lineskip=0pt\baselineskip=0pt
          \hbox{$<$}\hbox{$\sim$}}}}
\def\gsim{\mathrel{\lower2.5pt\vbox{\lineskip=0pt\baselineskip=0pt
          \hbox{$>$}\hbox{$\sim$}}}}

\begin{document}   
\markright{Scalar fields, energy conditions, and traversable wormholes \hfil}
\def\Barcelo{Barcel\'o}
\def\Paris{Par\'\i{}s}
\title{\Large \bf Scalar fields, energy conditions, \\
and traversable wormholes}
\author{Carlos \Barcelo\ and Matt Visser\\[2mm]
{\small \it 
Physics Department, Washington University, 
Saint~Louis, Missouri 63130-4899, USA.}}
\date{{\small 7 March 2000; \LaTeX-ed \today}}
\maketitle
\begin{abstract}
We describe the different possibilities that a simple and apparently
quite harmless classical scalar field theory provides to violate the
energy conditions. We demonstrate that a non-minimally coupled scalar
field with a positive curvature coupling $\xi>0$ can easily violate
all the standard energy conditions, up to and including the averaged
null energy condition (ANEC).  Indeed this violation of the ANEC
suggests the possible existence of traversable wormholes supported by
non-minimally coupled scalars.  To investigate this possibility we
derive the classical solutions for gravity plus a general (arbitrary
$\xi$) massless non-minimally coupled scalar field, restricting
attention to the static and spherically symmetric configurations.
Among these classical solutions we find an entire branch of
traversable wormholes for every $\xi>0$. (This includes and
generalizes the case of conformal coupling $\xi=1/6$ we considered in
Phys. Lett. {\bf B466} (1999) 127--134.)  For these traversable
wormholes to exist we demonstrate that the scalar field must reach
trans-Planckian values somewhere in the geometry.  We discuss how this
can be accommodated within the current state of the art regarding
scalar fields in modern theoretical physics. We emphasize that these
scalar field theories, and the traversable wormhole solutions we
derive, are compatible with all known experimental constraints from
both particle physics and gravity.

\vspace*{5mm}
\end{abstract}
\vfill
\hrule
\bigskip
\centerline{\underline{E-mail:} {\sf carlos@hbar.wustl.edu}}
\centerline{\underline{E-mail:} {\sf visser@kiwi.wustl.edu}}
\bigskip
\centerline{\underline{Homepage:} {\sf http://www.physics.wustl.edu/\~{}carlos}}
\centerline{\underline{Homepage:} {\sf http://www.physics.wustl.edu/\~{}visser}}
\bigskip
\centerline{\underline{Archive:}
{\sf gr-qc/0003025}}
\bigskip
\hrule
\clearpage
\def\Box{\nabla^2}
\def\d{{\mathrm d}}
\def\ie{{\em i.e.\/}}
\def\eg{{\em e.g.\/}}
\def\etc{{\em etc.\/}}
\def\etal{{\em et al.\/}}
\def\S{{\mathcal S}}
\def\I{{\mathcal I}}
\def\L{{\mathcal L}}
\def\R{{\mathcal R}}
\def\eff{{\mathrm{eff}}}
\def\Newton{{\mathrm{Newton}}}
\def\bulk{{\mathrm{bulk}}}
\def\matter{{\mathrm{matter}}}
\def\tr{{\mathrm{tr}}}
\def\normal{{\mathrm{normal}}}
\def\implies{\Rightarrow}
\def\half{{1\over2}}
\def\SIZE{1.00} 
\section{Introduction}
\label{S:introduction}

It is often (mistakenly) believed that every kind of matter, on scales
in which we do not need to consider its quantum features, has an
energy density that is everywhere positive.  (In fact, we could think
on this property as defining what we understand by classical matter).
This is more precisely stated by saying that every type of classical
matter satisfies the energy conditions of general
relativity~\cite{hawking}.  There are several (pointwise) energy
conditions requiring that various linear combinations of the
components of the energy-momentum tensor of matter have positive
values. (Or at the very least, non-negative values.) Among them, we
will be particularly interested in the null energy condition (NEC)
because it is the weakest one: If the NEC is violated all pointwise
energy conditions would be violated~\cite{visser1}.

Now, if we assume that the energy conditions are satisfied by every
kind of classical matter, general relativity leads to many powerful
classical theorems. The singularity theorems~\cite{hawking}, the
positive mass theorem~\cite{schoen}, the superluminal censorship
theorem~\cite{olum,visser2,visser2a}, the topological censorship
theorem~\cite{friedman}, certain types of no-hair
theorem~\cite{bekenstein-no-hair}, and various constraints on black
hole surface gravity~\cite{dbh}, all make use of some type of energy
condition.  As an illustration of the importance of these results, the
conclusion regarding the inevitability of the appearance of
singularities~\cite{hawking}, has been (for the last thirty years) one
of the central pillars from which many investigations in general
relativity start.
(Additional powerful technical assumptions are also needed for the
singularity theorems to apply; see~\cite{senovilla} for a critical
review).
In this regard, the alleged impossibility of the existence of
traversable wormholes connecting different spatial regions of the
universe~\cite{friedman,morris1}, a topic in which we will be
particularly interested in this paper, could be seen as a
complementary way of phrasing that conclusion (of course, one should
not carry this parallelism too far).

Assuming the positivity of the energy density implies that spacetime
geometries containing traversable wormholes\footnote{
The term ``traversable wormhole'' was adopted by M. Morris and 
K.S.~Thorne~\cite{morris1} to describe a class of Lorentzian 
geometries connecting two asymptotically flat regions of 
spacetime, in a manner suitable for a signal or particle
to pass through in both direcctions. By the definition of 
traversability it should be possible to travel from one asymptotic 
region to the other without encountering either horizons or 
naked singularities. In this paper we use this term exclusively
form this geometrical point of view.}
are ruled out of the
classical realm. Specifically, the topological censorship
theorem~\cite{friedman} states that if the averaged null energy (ANEC;
the NEC averaged over a complete null geodesic) is satisfied, then
there cannot be any topological obstruction (\eg, a wormhole) inside
any asymptotically flat spacetime.  In agreement with this result, a
purely local analysis by David Hochberg and one of the present
authors~\cite{hochberg1} shows that the violation of NEC on or near
the throat of a traversable wormhole is a generic property of these
objects.  For this reason most of the investigations regarding
traversable wormholes tend to view these objects as semiclassical in
nature, using the expectation value of the quantum operator associated
with the energy-momentum tensor as the source of
gravity~\cite{hochberg2}.

However, it is easy to demonstrate that an extremely simple and
apparently quite innocuous classical field theory, a scalar field
non-minimally coupled to gravity (that is, with a non-vanishing
coupling to the scalar curvature), can violate the NEC and even the
ANEC~\cite{flanagan,barcelo,visser3}.  As we will see, some of the
other energy conditions can be violated even by minimally coupled
scalar fields.  There exist other classical systems that exhibit NEC
violations, such as Brans--Dicke
theory~\cite{agnese1,nandi,anchordoqui,hochberg3}, higher derivative
gravity~\cite{hochberg4} or Gauss--Bonnet theory~\cite{bhawal}, but
they are all based on modifications of general relativity at high
energies.  It is the simplicity of the scalar field theory that
particularly attracted our attention.

In a previous paper~\cite{barcelo} we analyzed a massless scalar field
conformally coupled to gravity (that is, we used the special value
$\xi=1/6$) and found that among the classical solutions for the system
there exists an entire branch of traversable wormholes.  The case of
conformal coupling has many interesting features and, in fact, it
seems the most natural behaviour for a scalar field at low
energies~\cite{barcelo,visser3}.  However, in this paper we want to
point out that assuming conformal coupling is not the critical issue,
and that it is only the fact of non-minimality (more precisely,
positive curvature coupling $\xi>0$) that is important for traversable
wormhole solutions to exist.  We have found general expressions for
the classical solutions of gravity plus a massless non-minimally
coupled scalar field.  In particular, we have found a sub-class of
traversable wormhole solutions for the entire range $\xi> 0$.
In our solutions, apart from the geometry, there is also a scalar
field that modifies the effective Newton constant. We will not
address here the possible effects of this scalar field on the 
journey of any hypothetical traveller trying to cross the wormhole
throat.

In the next section we will review the possibilities that scalar
fields offer to violate the different energy conditions. (That scalar
fields might potentially cause problems in this regard was first noted
at least 25 years ago~\cite{bekenstein-bounce}, and is an issue that
has periodically come in and out of focus since then~\cite{visser3}.)
Then we will obtain and describe the classical solutions for gravity
plus a non-minimally coupled massless scalar field, restricting to
static and spherically symmetric configurations.  Among these
solutions there are an assortment of naked singularities but we also
find an entire branch of traversable wormhole solutions for curvature
coupling $\xi>0$. We will leave it to section \ref{S:wormhole} to
analyze these traversable wormhole geometries in detail.  Finally, in
section \ref{S:discussion} we will discuss the key features of these
solutions and the plausibility with which they might actually be
produced in nature, based on the role of scalar fields in modern
theoretical physics and the totality of the experimental constraints
arising from both particle physics and gravity physics.

\section{Scalar fields and energy conditions}
\label{S:conditions}

When a classical scalar field acts as a source of gravity, many of the
energy conditions can be violated depending on the form of the scalar
potential and the value of the curvature coupling.

\subsection{Effective stress energy tensor}

The Einstein equations $\kappa \; G_{\mu\nu} = T_{\mu\nu}$ relate the
geometry-dependent Einstein tensor $G_{\mu\nu}$ to the
energy-momentum tensor for the matter field $T_{\mu\nu}$. The symbol
$\kappa$ represents essentially the inverse of Newton's constant,
$\kappa =1/(8 \pi G_N)$.  These equations can be obtained by varying
the Einstein--Hilbert action which, for a generically
coupled scalar field, reads
\begin{equation}
\S
=
{1 \over 2}\int \;d^4x\sqrt{-g}\; \kappa \; R
+\int \;d^4x\sqrt{-g} 
\left(
-{1 \over 2} g^{\mu\nu} \; \partial_{\mu}\phi_\xi \; \partial_{\nu}\phi_\xi 
-V(\phi_\xi)-{1 \over 2}\xi \;R \;\phi_\xi^2 
\right).
\label{E:nlagrangian}
\end{equation}
Then, the scalar field energy-momentum tensor has the
form~\cite{flanagan}
\begin{eqnarray}
[T(\phi_\xi)]_{\mu\nu} 
=&&\hspace*{-6mm} 
\nabla_{\mu} \phi_\xi \; \nabla_{\nu} \phi_\xi 
- {1 \over 2} g_{\mu\nu} (\nabla \phi_\xi)^2 
- g_{\mu\nu} \; V(\phi_\xi) 
\nonumber \\
&&\hspace*{-6mm}
+\xi\left[ G_{\mu\nu} \; \phi_\xi^2 
- 2 \;\nabla_{\mu} ( \phi_\xi \; \nabla_{\nu} \phi_\xi) 
+ 2 \; g_{\mu\nu} \; \nabla^{\lambda} (\phi_\xi \; \nabla_{\lambda} \phi_\xi) 
\right].
\label{E:emt}
\end{eqnarray}
This energy-momentum tensor has a term that depends algebraically on
the Einstein tensor.  By grouping all the dependence on $G_{\mu\nu}$
on the left hand side of Einstein equations we can rewrite them,
alternatively, by using an effective energy-momentum tensor
\begin{eqnarray}
[T^{\eff}(\phi_\xi)]_{\mu\nu} 
=&&\hspace*{-6mm}
{\kappa \over \kappa-  \xi \phi_\xi^2}
\bigg[ 
\nabla_{\mu} \phi_\xi \; \nabla_{\nu} \phi_\xi 
- {1 \over 2} g_{\mu\nu} (\nabla \phi_\xi)^2 
- g_{\mu\nu} \; V(\phi_\xi) 
\nonumber \\
&&\hspace*{-6mm}
-\xi\left[ 
2 \;\nabla_{\mu} ( \phi_\xi \; \nabla_{\nu} \phi_\xi) 
- 2 \; g_{\mu\nu} \; \nabla^{\lambda} (\phi_\xi \; \nabla_{\lambda} \phi_\xi) 
\right]
\bigg].
\label{E:emt2}
\end{eqnarray}
This is the relevant expression for the analysis of the different
energy conditions: Since the Einstein equations now read $\kappa \;
G_{\mu\nu} = [T^\eff(\phi_\xi)]_{\mu\nu}$, a constraint on this
effective stress-energy tensor is translated directly into a
constraint on the spacetime curvature, and it is ultimately
constraints on the spacetime curvature that lead to singularity
theorems and the like.

\subsection{Pointwise energy conditions}

Now let $v^{\mu}$ be a properly normalized timelike vector,
$(v^2=-1)$, and, for convenience, let it be locally extended to a
geodesic vector field, so that $v^{\mu}\nabla_{\mu}v^{\nu} = 0$. If
$x^{\mu}(\tau)$ denotes a timelike geodesic with tangent vector
$v^{\mu} = \d x^{\mu}/\d\tau$ we have $v^{\mu}\nabla_{\mu}\phi =
\d\phi/\d\tau \equiv \phi'$.  We can now express the strong energy
condition (SEC) as
\begin{eqnarray}
 R_{\mu\nu} \; v^{\mu} \; v^{\nu}
=&&\hspace*{-6mm}
{1 \over \kappa}
\left(
[T^\eff(\phi_\xi)]_{\mu\nu}
- {1 \over 2}g_{\mu\nu} \; [T^\eff(\phi_\xi)]\,\right) 
v^{\mu} \; v^{\nu}
\nonumber \\
=&&\hspace*{-6mm}
{1 \over \kappa- \xi \phi_\xi^2} 
\left[ 
(\phi_\xi ')^2-V(\phi_\xi)-\xi[(\phi_\xi^2)''-\nabla^\mu
(\phi_\xi\nabla_\mu\phi_\xi)]
\right]\geq 0,
\label{E:sec}
\end{eqnarray}
where $R_{\mu\nu}$ is the Ricci tensor of the geometry.  It is easy to
see that even in the minimally coupled case the SEC can be violated
for positive values of the scalar potential such us a mass term or a
positive cosmological constant.  Every cosmological inflationary
process violates the SEC~\cite{visser3,science}.  We could say that
violation of the SEC is a generic property of scalar fields. In fact,
it is so easy to violate SEC in many situations that it has almost
become to be abandoned~\cite{visser3} as a reasonable restriction on
the properties of matter.

With the same definitions, the weak energy condition (WEC) reads
\begin{eqnarray}
G_{\mu\nu} \; v^{\mu} \; v^{\nu}
=&&\hspace*{-6mm}
{1 \over \kappa}\;
[T^\eff(\phi_\xi)]_{\mu\nu} \; v^{\mu} \; v^{\nu}
\\
=&&\hspace*{-6mm}
{1 \over \kappa- \xi \phi_\xi^2} 
\left[ 
(1-2\xi)(\phi_\xi')^2 
+{1 \over 2}(\nabla \phi_\xi)^2 +V(\phi_\xi)
-\xi[(\phi_\xi^2)''
+2\nabla^\mu(\phi_\xi \nabla_\mu\phi_\xi)]
\right] \geq 0.
\nonumber
\label{E:wec}
\end{eqnarray}
In the minimally coupled case $(\xi=0)$ it is generically satisfied.
Only a large negative potential, for example, a negative cosmological
constant, could provide a violation of WEC.  Although recent
observations suggest a probable positive value for the effective
cosmological constant, a possible negative value can not be ruled out
on theoretical grounds.  In the non-minimal case there are various
terms that can be negative depending on the situation, so the WEC can
be violated in various ways.

Finally, let us now analyze the NEC.  It is the weakest pointwise
energy condition, that is, when it is violated the WEC and SEC are
violated too.  Let $k^{\mu}$ be a null vector tangent to the null
geodesic $x^{\mu}(\lambda)$, with $\lambda$ some affine parameter.  In
an analogous way as with the previous energy conditions, we arrive at
the following expression for the NEC
\begin{equation}
G_{\mu\nu} \; k^{\mu} \; k^{\nu}
=
{1 \over \kappa}\;
[T^\eff(\phi_\xi)]_{\mu\nu} \; k^{\mu} \; k^{\nu}
=
{1 \over  \kappa- \xi\phi_\xi^2}
\left[\phi_\xi'^2 - \xi(\phi_\xi^2)''\right] \geq 0.
\label{E:nec}
\end{equation}
This condition is clearly satisfied by minimally coupled scalars.
However, for $\xi\neq 0$ it can be violated in a number of ways: For
$\xi < 0$ any local minimum of $\phi_\xi^2$ violates the NEC while for
$\xi > 0$ and $|\phi_\xi|$ small, [meaning $|\phi_\xi| < (\kappa/
\xi)^{1/2}$], any local maximum of $\phi_\xi^2$ violates the NEC. Finally
for $\xi > 0$ and $|\phi_\xi|$ large, [meaning $|\phi_\xi| > (\kappa/
\xi)^{1/2}$, roughly corresponding to super--Planckian values for the
scalar field], any local minimum of $\phi_\xi^2$ violates the NEC.

At this point it is worth noticing that, because our analysis is
completely classical, it is {\em a priori} conceivable that averaged
versions of the energy conditions (averaged over a geodesic) could in
principle be as easily violated as their pointwise counterparts.  In
particular, ANEC violations, critical for traversable wormhole
configurations to be able to exist, could in principle be as easy to
find as NEC violations, and it is to exploring this possibility that
we now turn.

\subsection{Averaged energy conditions --- ANEC}

Suppose we take a segment of a null geodesic and consider the ANEC
integral
\begin{equation}
\I(\lambda_1,\lambda_2) = 
\int_{\lambda_1}^{\lambda_2} 
[T^\eff(\phi_\xi)]_{\mu\nu} \; k^\mu \; k^\nu \; \d\lambda.
\end{equation}
Then
\begin{equation}
\I(\lambda_1,\lambda_2) = 
\int_{\lambda_1}^{\lambda_2} 
{\kappa\over\kappa-\xi\phi_\xi^2} 
\left\{
\left({\d\phi_\xi \over \d\lambda}\right)^2 -
2 \xi
{\d\over \d\lambda} \left( \phi_\xi {\d\phi_\xi\over \d\lambda} \right)
\right\} \d\lambda.
\end{equation}
Integrate by parts
\begin{equation}
\I(\lambda_1,\lambda_2) = 
\int_{\lambda_1}^{\lambda_2} 
{\kappa\over\kappa-\xi\phi_\xi^2} 
\left\{
\left({\d\phi_\xi \over \d\lambda}\right)^2  +
{
4\xi^2 \phi_\xi^2  \; (\d\phi_\xi/\d\lambda)^2
\over
\kappa-\xi\phi_\xi^2
}
\right\} \d\lambda
- 
\left.\left\{
{2\xi\kappa\over\kappa-\xi\phi_\xi^2}
 \left( \phi_\xi \; {\d\phi_\xi\over \d\lambda} \right)
\right\}
\right|_{\lambda_1}^{\lambda_2} 
\end{equation}
Now assemble the pieces:
\begin{equation}
\I(\lambda_1,\lambda_2) = 
\int_{\lambda_1}^{\lambda_2} 
{\kappa [\kappa - \xi(1-4\xi)\phi_\xi^2 ]\over(\kappa-\xi\phi_\xi^2)^2} 
\left({\d\phi_\xi \over \d\lambda}\right)^2 
d\lambda
- 
\left.\left\{
{2\xi\kappa\over\kappa-\xi\phi_\xi^2}
 \left( \phi_\xi {\d\phi_\xi\over \d\lambda} \right)
\right\}
\right|_{\lambda_1}^{\lambda_2}. 
\end{equation}
Discarding the boundary terms is an issue fraught with subtlety: We
start by considering a complete null geodesic and assuming
sufficiently smooth asymptotic behaviour. Then the boundary terms from
asymptotic infinity can be neglected, and the only potential problems
with the boundary terms come from the places $\lambda_i$ where $\kappa
= \xi \phi_\xi(\lambda_i)^2$. We obtain
\begin{equation}
\I(-\infty,+\infty) = 
\oint 
{\kappa [\kappa - \xi(1-4\xi)\phi_\xi^2 ]\over(\kappa-\xi\phi_\xi^2)^2} 
\left({\d\phi_\xi \over \d\lambda}\right)^2 
d\lambda
+
\sum_i 
\left.\left\{
{2\xi\kappa\over\kappa-\xi\phi_\xi^2}
 \left( \phi_\xi {\d\phi_\xi\over \d\lambda} \right)
\right\}
\right|_{\lambda_i^+}^{\lambda_i^-}.
\end{equation}
\begin{enumerate}
\item
If $\xi<0$ then there are no places on the geodesic where $\kappa =
\xi [\phi_\xi(\lambda)]^2$, so the boundary terms represent an empty
set. But if $\xi<0$ the {\em integrand} in the above formula is itself
guaranteed positive so ANEC is satisfied.
\item
If $\xi>0$, but we have $(\phi_\xi)^2 < \kappa/\xi$, then again there
are no places on the geodesic where $\kappa = \xi
[\phi_\xi(\lambda)]^2$. Furthermore the {\em integrand} appearing
above is again positive and ANEC is satisfied.
\item
Finally, if $\xi>0$, and we have at least some places where
$(\phi_\xi)^2 > \kappa/\xi$, then there are by definition places on the
geodesic where $\kappa = \xi [\phi_\xi(\lambda)]^2$. The boundary terms
can no longer be neglected and potentially can contribute ---
typically making negative and infinite contributions to the ANEC
integral. Furthermore the {\em integrand} appearing above is no longer
guaranteed to be positive. (If $\xi \in (0, 1/4) $ then it is possible
to have $(\phi_\xi)^2 > \kappa / [\xi(1-4\xi)]$ and so make the
integrand negative.) In short: there is definitely the {\em
possibility} of ANEC violations under these conditions, and in the
exact solutions we investigate below we shall see that for some of
these exact solutions the ANEC is certainly violated.
\end{enumerate}

Thus we have seen that a rather simple and apparently quite harmless
scalar field theory can in many cases violate all the energy
conditions. Violating all the pointwise energy conditions is
particularly simple, and violating the averaged energy conditions,
though more difficult, is still generically possible. In the next
section we shall exhibit some specific examples of this phenomenon in
the form of exact solutions to the coupled Einstein--scalar-field
equations.  In the final section we will discuss the extent to which
these exact solutions are realistic: we shall discuss the generic role
played by scalar fields in modern theoretical physics, and the
experimental/observational limits on their existence and behaviour in
order to address the physical plausibility of these energy condition
violations.

\section{Non-minimal classical solutions}
\label{S:solutions}

As we have just seen, a non-minimally coupled scalar field can
violate, in some circumstances, the NEC and even the ANEC. This opens
up the possibility of finding some traversable wormholes among the
many geometries that can be supported by a classical non-minimally
coupled scalar field.  That this is in fact the case for a massless
conformally coupled scalar field was shown in~\cite{barcelo}.  In the
present paper, we will see that even for non-conformal coupling
(curvature coupling different from $1/6$), but positive, we can also
find traversable wormholes.  In this section we will obtain the
classical solutions for gravity plus a generic non-minimally coupled
scalar field. For simplicity, we will restrict to the spherically
symmetric and static configuration and will take the scalar potential
$V(\phi)$ equal zero. (To add a particle mass to the scalar field
complicates the equations sufficiently to preclude the possibility of
analytic results).

\subsection{Some ``trivial'' solutions}

The first thing that we realize is that for any spatially constant
value of the scalar field, $\phi_\xi=C$, the Einstein equations reduce
to $\kappa \; G_{\mu\nu}=\xi \; G_{\mu\nu} \; C^2$.  By looking also
at the scalar field equation $(\nabla^2\phi_\xi = \xi\; R \; \phi_\xi)
\implies \xi\; R \; \phi_\xi = 0$, where $R$ is the Ricci scalar curvature,
we can straightforwardly find a variety of trivial solutions for this
system.  For (i) $\xi>0$ and $C\neq \pm \sqrt{\kappa / \xi}$, or (ii)
$\xi<0$ and any value value of $C$, we find that the ordinary vacuum
Einstein solutions also solve the coupled Einstein-scalar equations.
As we are here restricted to spherically symmetric and static
configurations without cosmological constant, the solutions that show
up are the Schwarzschild and anti-Schwarzschild geometries ($M<0$
represents a perfectly good solution to the Einstein field equations,
normally the negative mass Schwarzschild geometry is excluded by hand,
here it's best to keep it for the time being as an aid in classifying
the total solution space.)

More surprising is that for $\xi>0$ and $C=\pm \sqrt{\kappa/\xi}$
every Ricci-scalar-flat geometry is a solution.  (That is, any
geometry satisfying the condition $R=0$ is a solution of the coupled
Einstein-scalar equations.)  The condition $R=0$ is characteristic of
geometries supported by conformally invariant matter. In particular,
the geometries that appear in the solutions for the $\xi=1/6$ case all
satisfy $R=0$.  These geometries were found in
Froyland~\cite{froyland} and also in~\cite{barcelo}, and will be
re-obtained later on this paper as particular cases. Here we want to
point out that these geometries are not only associated with the
conformal coupling but they appear quite generally, for arbitrary
$\xi>0$, provided only $\phi_\xi = C \equiv \pm \sqrt{\kappa/\xi}$.

\subsection{Solution generating technique}

To obtain additional (non-trivial) solutions for a scalar field
non-minimally coupled to gravity we will use a ``solution generating
technique'' that relies on knowledge of the solutions for the minimally
coupled case~\cite{bekenstein,duruisseau}.  The classical solutions
for a massless scalar field minimally coupled to gravity are very well
known.  They have been discovered and re-discovered several times in
different coordinate systems (see the articles by
Fisher~\cite{fisher}, Janis, Newman, and Winicour~\cite{jnw},
Wyman~\cite{wyman}, and M.~Cavaglia and V.~De Alfaro~\cite{cavaglia}).  
They can be expressed~\cite{agnese2} as
\begin{eqnarray}  
&&\hspace*{-8mm}
ds_m^2
=
-\left(1-{2\eta \over r} \right)^{\cos\chi} dt^2
+\left(1-{2\eta \over r} \right)^{-\cos\chi} dr^2
+\left(1-{2\eta \over r} \right)^{1-\cos\chi} 
r^2 (d\theta^2+\sin^2 \theta \; d\Phi^2), 
\nonumber  \\ 
&&          \\
&&\hspace*{+3.6cm}
\phi_m
=
\sqrt{{\kappa \over 2 } }\; \sin \chi \;  
\ln \left(1-{2\eta \over r} \right).
\label{E:msolution}
\end{eqnarray}
The same geometry, which has an obvious symmetry under $\chi\to
-\chi$, can exist with a field configuration $\phi_m$ or the reversed
sign configuration $-\phi_m$.  Less obvious is that by making a
coordinate transformation $r\to\tilde r = r - 2\eta$, one uncovers an
additional symmetry under $\{ \eta,\chi \} \to \{ -\eta,\chi+\pi \}$,
(with $\phi_m \to +\phi_m$).  The key to this symmetry is to realize
that
\begin{equation}
\left( 1 - {2\eta\over r} \right) = 
\left( 1 + {2\eta\over \tilde r} \right)^{-1}.
\end{equation}
In view of these symmetries one can without loss of generality take
$\eta \geq 0$ and $\chi \in [0, \pi]$ remembering the overall two
possible signs for the scalar field.  Similar symmetries will be
encountered for non-minimally coupled scalars.

The Lagrangian for which these ``minimal'' solutions are extrema is 
\begin{equation}
S_m={1 \over 2}\int \;d^4x\sqrt{-g_m}\; \kappa \; R_m
+\int \;d^4x\sqrt{-g_m} 
\left(-{1 \over 2} g_m^{\mu\nu} \; 
\partial_{\mu}\phi_m \; \partial_{\nu}\phi_m \right).
\label{E:mlagrangian}
\end{equation}
This Lagrangian and the Lagrangian for a non-minimally coupled
massless scalar field,
\begin{equation}
S_\xi={1 \over 2}\int \;d^4x\sqrt{-g_\xi}\; \kappa \; R_\xi
+\int \;d^4x\sqrt{-g_\xi} 
\left(-{1 \over 2}g^{\mu\nu}_{\xi}\; 
\partial_{\mu}\phi_\xi \; \partial_{\nu}\phi_\xi 
-{1 \over 2}\xi\;R_\xi \; \phi_\xi^2 \right),
\label{E:nlagrangian2}
\end{equation}
can be related by a conformal transformation of the metric
$g_{\mu\nu}^\xi=\Omega^2 \; g_{\mu\nu}^m$ and a redefinition of the
scalar field $\phi_\xi=\phi_\xi(\phi_m)$ \cite{bekenstein}.  Rescaling
the field $\Phi=\phi/\sqrt{6\kappa}$ we can write the specific
transformation as
\begin{equation}
\Omega^2={1 \over (1-6\xi\;\Phi_\xi^2)}, 
\hspace{10mm} 
{d \Phi_m \over d\Phi_\xi}=\pm {\sqrt{1-6\xi(1-6\xi)\; \Phi_\xi^2} \over
(1-6\xi\;\Phi_\xi^2)}.
\label{E:equations}
\end{equation}
Notice that for $0<\xi<1/6$ the absolute value of the non-minimally
coupled scalar field cannot surpass $1 / \sqrt{6\xi(1-6\xi)}$ if we
want expression (\ref{E:equations}) to make sense.  For every solution of
equation (\ref{E:equations}) we have a two-parameter family, $\{\eta,
\chi\}$, of solutions for the non-minimally coupled system.  For
$\xi=0$, the expressions in (\ref{E:equations}) become the identity
transformation, as they must.  For $\xi=1/6$ we easily find the
following solutions:
\begin{eqnarray}  
&\Omega^2=\cosh^2(\Phi_m-\Phi^0_{+}),
\hspace*{2cm}
&\Phi_{\xi=1/6}=\pm\tanh(\Phi_m-\Phi^0_{+}),
\label{E:physicalsol1} 
\\ 
\nonumber
\\
&\Omega^2=-\sinh^2(\Phi_m-\Phi^0_{-}),
\hspace*{1.65cm}
&\Phi_{\xi=1/6}=\pm\coth(\Phi_m-\Phi^0_{-}),
\label{E:unphysicalsol}
\end{eqnarray}
with $\Phi^0_{+}$ and $\Phi^0_{-}$ two arbitrary real constants.  The
second set of solutions is unphysical because it gives a negative sign
for the metric signature (opposite to the one we are using).  However,
one can easily demonstrate~\cite{bekenstein} that the Einstein tensor
$G_{\mu\nu}$ and the energy-momentum tensor (\ref{E:emt}) for a
massless scalar field are both invariant if we change $g_{\mu\nu}$ to
$-g_{\mu\nu}$ leaving the field unchanged.  Therefore, from the
unphysical solutions (\ref{E:unphysicalsol}) we obtain physical
solutions of the form
\begin{equation} 
\Omega^2=\sinh^2(\Phi_m-\Phi^0_{-}),
\hspace*{2cm}
\Phi_{\xi=1/6}=\pm\coth(\Phi_m-\Phi^0_{-}).
\label{E:physicalsol2}
\end{equation}
We can also find additional solutions by considering the limiting
cases when the constants $\Phi^0_{+}$ or $\Phi^0_{-}$ tend to ($\pm$)
infinity.  In this way we find the solutions:
\begin{eqnarray}  
&\Omega^2={\rm exp}(2\Phi_m),
\hspace*{2cm}
&\Phi_{\xi=1/6}=\pm 1,
\label{E:physicalsol1l}  
\\ 
\nonumber
\\
&\Omega^2={\rm exp}(-2\Phi_m),
\hspace*{1.65cm}
&\Phi_{\xi=1/6}=\pm 1.
\label{E:physicalsol2l}
\end{eqnarray}
All these metrics $g_{\mu\nu}^{(\xi=1/6)}=\Omega^2 \; g_{\mu\nu}^m$,
described in (\ref{E:physicalsol1}), (\ref{E:physicalsol2}),
(\ref{E:physicalsol1l}), and (\ref{E:physicalsol2l}) have a zero
scalar curvature, $R=0$, owing to the conformal coupling
features~\cite{barcelo}.  Therefore, any of these geometries,
supplemented with the specific value $\Phi_\xi=\pm 1/\sqrt{6\xi}$ for
the non-minimally coupled scalar field (with $\xi>0$) are examples of
the ``trivial'' solutions for the non-minimally coupled system of
which we have spoken at the beginning of this section,\footnote{
Notice that, for the conformal case the solutions obtained by means of
the limiting procedure, (\ref{E:physicalsol1l}) and
(\ref{E:physicalsol2l}), are already in this later class.}
and these ``trivial'' solutions no longer seem to be all that trivial.
These solutions cannot be obtained by means of the conformal
transformation procedure described above because the conformal factor
has a singular behaviour for those scalar field values.

The spacetime geometries of all these solutions were analyzed in a
previous paper~\cite{barcelo} although with a slightly different
parameterization.  Here, we will describe them as limiting cases of
the solutions that we will find for the interval $0<\xi<1/6$. However
we want to mention that among these conformal solutions we found
traversable wormholes and that in the present parameterization they
correspond to (\ref{E:physicalsol1}) with $\Phi^0_{+}>0$,
(\ref{E:physicalsol2}) with $\Phi^0_{-}>0$, and
(\ref{E:physicalsol2l}); all using a value for $\chi$ in
(\ref{E:msolution}) equal to $\pi/3$.

Let us now solve the general equation (\ref{E:equations}) for 
an arbitrary $\xi$. We first rewrite this equation as a sum of two terms
\begin{equation}
{d \Phi_m \over d\Phi_\xi}
=\pm\left[ {6\xi \over (1-6\xi\Phi_\xi^2) \sqrt{1-6\xi(1-6\xi)\Phi_\xi^2}}
+{(1-6\xi) \over \sqrt{1-6\xi(1-6\xi)\Phi_\xi^2}}\right].
\label{E:requation}
\end{equation}
For convenience, we will express the solutions of this equation
in terms of two functions $F(\Phi_\xi)$ and $H(\Phi_\xi)$ as
\begin{equation}
\Phi_m(\Phi_\xi)=\pm\ln [F(\Phi_\xi) \; H(\Phi_\xi)]. 
\label{E:fh}
\end{equation}
The function $F$ will be related with the first term in (\ref{E:requation})
and the function $H$ with the second term.

\subsection{$F(\Phi)$:}

The first term in (\ref{E:requation}) can easily be integrated by 
changing to a new variable
\begin{equation}
v\equiv{ 6\xi\Phi_\xi \over \sqrt{1-6\xi(1-6\xi)\Phi_\xi^2}}.
\label{E:change}
\end{equation}
Using this new variable we only have to solve an integral of the form
\begin{equation}
\int \; {dv \over 1-v^2},
\label{E:integral}
\end{equation}
that we express in terms of logarithms.  In this way (once we invert the 
change of variables), we arrive at a closed form for the function F in
(\ref{E:fh}), for arbitrary $\xi$.  Indeed, the above integral gives
rise to two possible functions $F$, $F_{+}$ and $F_{-}$, of the form
\begin{equation}
F_{\pm}(\Phi_\xi)=\Phi_\pm 
\sqrt{\pm { \sqrt{1-6\xi(1-6\xi)\Phi_\xi^2}+6\xi\Phi_\xi \over
\sqrt{1-6\xi(1-6\xi)\Phi_\xi^2}-6\xi\Phi_\xi }}
\label{E:f}
\end{equation}
for $\xi>0$, and only one, the positive sign $F_{+}$, for $\xi\leq 0$.

This restriction on signs is due to the fact that for $\xi>0$ the
absolute value of the variable $v$ can be greater or lower than unity
and these two domains have to be analyzed separately. In terms of the
scalar field, these two domains are separated by the critical value
$|\Phi|=1/\sqrt{6\xi}$.  At best, the function $F_{+}$ is only defined
for $|\Phi| \leq 1/ \sqrt{6\xi}$ whilst $F_{-}$ is at best defined for
the on region $|\Phi| \geq 1/ \sqrt{6\xi}$. Indeed, in the case
$0<\xi<1/6$, the function $F_{-}$ is only defined up to $|\Phi| \leq 1
/\sqrt{6\xi(1-6\xi)}$.

An observation is in order at this point.  In the second domain of
values for the scalar field the conformal factor $\Omega^2=1/
(1-6\xi\Phi_\xi^2)$ is negative and so the geometry obtained is
``unphysical'' in the sense that the metric has reversed signature.
As explained before for the conformal case, from this unphysical
solution we can obtain a physical solution by changing the sign of the
conformal factor to $\Omega^2=1/(6\xi\Phi_\xi^2-1)$~\cite{bekenstein}.
Another observation concerning the functions $F_{\pm}$ is that we have
already embedded into them the corresponding integration constants
for each solution.  The $\Phi_\pm$ in (\ref{E:f}) are these
integration constants.  Owing to the logarithmic form in which we have
cast the solution these constants are both positive. In summary ---
\begin{enumerate}
\item[{\it\bf i)}] $\xi<0$:

$F_+(\Phi)$ is real and well-defined for all values of $\Phi$; 
\\
$F_-(\Phi)$ is undefined (complex), and un-needed.

\item[{\it\bf ii)}] $\xi=0$:

$F_+(\Phi) = \Phi_+$;
\\
$F_-(\Phi)$ is undefined (complex), and un-needed.

\item[{\it\bf iii)}] $0 < \xi < 1/6$:

$F_+(\Phi)$ is real and well-defined for 
$|\Phi|  < 1/\sqrt{6\xi} <  1/\sqrt{6\xi(1-6\xi)}$;
\\
and is undefined outside this range.
\\
$F_-(\Phi)$ is real and well-defined for 
$1/\sqrt{6\xi} <|\Phi| < 1/\sqrt{6\xi(1-6\xi)}$; 
\\
and is undefined outside this range.

\item[{\it\bf iv)}] $\xi = 1/6$:

For conformal coupling there is tremendous simplification
\[
F_{\pm}(\Phi)=\Phi_\pm \;
\sqrt{\pm { 1+\Phi \over 1-\Phi }}.
\]
$F_+(\Phi)$ is well defined for $|\Phi|<1$, whereas $F_-(\Phi)$ is
well-defined for $|\Phi| > 1$.

\item[{\it\bf v)}] $\xi > 1/6$:

$F_+(\Phi)$ is real and well-defined for $|\Phi| < 1/\sqrt{6\xi}$; 
\\
and is undefined outside this range.
\\
$F_-(\Phi)$ is real and well-defined for $|\Phi| > 1/\sqrt{6\xi}$; 
\\
and is undefined outside this range.
\end{enumerate}

\subsection{$H(\Phi)$:}

The second term in (\ref{E:requation}) can be integrated directly
yielding different rather complex algebraic expressions for the
function $H(\Phi_\xi)$ depending on the value of $\xi$ ---
\begin{enumerate}
\item[{\it\bf i)}] $\xi<0$:
\begin{equation}
H(\Phi_\xi)
=
\left(
\sqrt{ 6\xi(6\xi-1)} \; \Phi_\xi
+\sqrt{1-6\xi(1-6\xi)\Phi_\xi^2} 
\right)^{\sqrt{{6\xi-1 \over 6\xi}}}.
\end{equation}
\item[{\it\bf ii)}] $\xi=0$:
\begin{equation}
H(\Phi_\xi)={\rm exp}(\Phi_\xi).
\end{equation}
\noindent
\item[{\it\bf iii)}] $0<\xi<1/6$:
\begin{equation}
H(\Phi_\xi)
=
\exp\left(
\sqrt{{1-6\xi \over 6\xi}} \; \sin^{-1} 
\left(\sqrt{ 6\xi(1-6\xi)}\Phi_\xi\right)
\right).
\end{equation}
\noindent
\item[{\it\bf  iv)}] $\xi=1/6$:
\begin{equation}
H(\Phi_\xi)=1.
\end{equation}
\noindent
\item[{\it\bf v)}] $\xi>1/6$:
\begin{equation}
H(\Phi_\xi)
=
\left(
\sqrt{ 6\xi(6\xi-1)} \; \Phi_\xi
+\sqrt{1-6\xi(1-6\xi)\Phi_\xi^2} 
\right)^{-\sqrt{{6\xi-1 \over 6\xi}}}.
\end{equation}
\end{enumerate}
Here, we have not introduced any arbitrary integration constants
because, as we mentioned before, we have already included these
constants in the $F_{\pm}$ functions.

\subsection{The general solution}

At this point we already have (implicitly) all the different classical
solutions for gravity plus a massless non-minimal scalar field. On one
hand, if we substitute the different $F(\Phi_\xi)$ and $H(\Phi_\xi)$
in the left hand side of (\ref{E:fh}), and the minimally coupled
scalar field (\ref{E:msolution}) in the right hand side, we have a
implicit expression for the non-minimally coupled scalar field as a
function of the radial coordinate
\begin{equation}
\left(1-{2\eta \over r}\right)^{{\sin\chi \over 2 \sqrt{3}}}=
[F(\Phi_\xi)H(\Phi_\xi)]^{\pm 1}.
\label{E:fieldr}
\end{equation}
These relations cannot be analytically inverted in general.  Also,
depending on whether the function $F$ is $F_{+}$ or $F_{-}$ we have
geometries $g^{\xi}_{\mu\nu}=\Omega^2 \; g^{m}_{\mu\nu}$ with
different conformal factors:
\begin{equation}
F_{+} \rightarrow \Omega^2={1 \over  1-6\xi\; \Phi_\xi^2};
\hspace*{15mm}
F_{-} \rightarrow \Omega^2={1 \over 6\xi\; \Phi_\xi^2-1 }.
\end{equation}

Our next step is to analyze the different solutions found.  Let us
begin with some general comments.  For convenience, henceforth we will
work in isotropic coordinates,
\begin{equation}
r={\bar r}\left(1+{\eta \over 2{\bar r}}\right)^2.
\label{E:isotropic}
\end{equation}
In these coordinates the previous implicit expression for 
the scalar field (\ref{E:fieldr}) becomes
\begin{equation}
\left[ {1- {\eta \over 2 {\bar r}} \over 
1+ {\eta \over 2 {\bar r}}} \right]^{ {\sin\chi \over \sqrt{3} }}=
[F(\Phi_\xi)H(\Phi_\xi)]^{\pm 1}.
\label{E:implicitrbar}
\end{equation}
As in the minimally coupled case, the non-minimal solutions possess a
symmetry under $(\eta, \chi)$ going to $(-\eta, \chi+\pi)$. Also, they
exhibit a symmetry under the change of $\chi$ to $-\chi$ and a
simultaneous flip in the sign of the exponent on the right hand side
of (\ref{E:implicitrbar}).  Therefore, we will without loss of
generality restrict the analysis to $\eta\geq 0$, $\chi\in[0,\pi]$ and
the positive exponent, remembering that for every solution there is a
second solution that is geometrically identical but with a reversed
sign for the scalar field.

After a little algebra we can express the Schwarzschild radial
coordinate $\R$ for the different geometries as a function of
$\Phi_\xi$ and, therefore, implicitly as a function of the isotropic
radial coordinate ${\bar r}$,
\begin{equation}
\R_\pm(\Phi_\xi)=2\eta \;
{ (F_{\pm} H)^{{\sqrt{3}(1-\cos\chi) \over  \sin\chi}} \over 
\left[1 -(F_{\pm} H)^{{2\sqrt{3} \over  \sin\chi}}\right]
\sqrt{\pm (1-6\xi\Phi_\xi^2)}}.
\label{E:src}
\end{equation}
In the following discussion it is useful to know that, in the case
$\xi>0$, if we make a perturbative expansion of the functions
$F_{\pm}H$ around $\Phi_\xi =-1/\sqrt{6\xi}$ (that is, we take
\begin{equation}
\Phi_\xi= {-1 \over \sqrt{6\xi}}\pm\epsilon,
\end{equation}
with $\epsilon$ and small positive quantity) then the functions
$F_{\pm}H\rightarrow 0$ as $\sqrt{\epsilon}$.  In the same way, the
conformal factor $\Omega\rightarrow \infty $ as $1/\sqrt{\epsilon}$.

\subsection{Solutions with $\xi<0$}

In this case, the function $F_{+}H$ is everywhere positive. (See figure
1.)  For a certain (finite) value of the scalar field (depending on
$\Phi_+$ and $\xi$), $F_{+}H \rightarrow 1 $.  We can easily verify
that both ${\bar r}$ and ${\cal R}_{+}$ go to infinity at this stage,
so they are describing an asymptotic region. Indeed, it is an
asymptotically flat region because since the scalar field goes to a
finite constant in the asymptotic region with ${\bar r} \rightarrow
\infty$ the conformal factor tends to a constant, and therefore the
geometry behaves as in the minimal solution metric
(\ref{E:msolution}).

Decreasing the value of the scalar field we drive ourselves towards
the interior of the geometry.  As $\Phi_\xi \rightarrow -\infty$,
$F_{+}H \rightarrow 0$, and ${\bar r} \rightarrow \eta/2$. Then,
analyzing the asymptotic behavior of $\Phi_\xi$ in equation
(\ref{E:src}) we can conclude that the Schwarzschild radial coordinate
shrinks to zero for every $\chi\neq 0$.  Thus, for a non-minimally
coupled scalar field with a curvature coupling $\xi<0$ we find naked
singularities, in the same manner as for a minimally coupled scalar
field~\cite{wyman}.  For the special case $\chi=0$ we find solutions
with a constant value for the scalar field and a Schwarzschild
geometry (of course, in this case the spacetime geometry does not
shrink to zero for ${\bar r}=\eta/2$), while for $\chi=\pi$ we
encounter an anti-Schwarzschild geometry.

\subsection{Solutions with $\xi=0$}

The function $F(\Phi)$ in this case becomes a constant. The function
$H(\Phi)$ is ${\rm exp}(\Phi_\xi)$, so in equation (\ref{E:fh}) we can
read that $\Phi_m=\Phi_\xi+const$. Clearly, we recover the standard
minimally coupled solutions with its naked singularities.

\subsection{Solutions with $0<\xi<1/6$}
 
Here we have to analyze separately both signs in equation
(\ref{E:src}).  Let us begin with the positive, $F_{+}$.  As before,
for a certain value of the scalar field the function $F_{+}H
\rightarrow 1 $, making both $\R_{+}\rightarrow \infty$ and
${\bar r} \rightarrow \infty$, thereby describing an asymptotically
flat region. (See figure 2.)  Then, decreasing the value of the field
we leave the asymptotic region going towards the interior of the
geometry. At the value $-1/\sqrt{6\xi}$ the ${\bar r}$ coordinate
reaches the value $\eta/2$ with the corresponding zero value for
$F_{+}H$. A perturbative analysis around $-1/\sqrt{6\xi}$ tells us
that for $\chi\in(0, \pi/3)$ the Schwarzschild radial coordinate blows
up. For the rest of values the geometry shrinks to a naked singularity
except for $\chi=0$ and $\chi=\pi/3$ in which the Schwarzschild radial
coordinate goes to a constant value.

The solution with $\chi=0$ is once more the Schwarzschild geometry,
whilst among the naked singularities we have the $\chi=-\pi$ case
representing the anti-Schwarzschild geometry.

The solutions with $\chi=\pi/3$ are more bizarre. At ${\bar r}=\eta/2$
the geometry neither shrinks to zero nor blows up to infinity.  We can
see the reason for this behaviour easily by looking at equation
(\ref{E:src}). For $\chi=\pi/3$ the exponent of the function $F_{+}H$
becomes unity and so it goes to zero at the same rate as the factor of
$\sqrt{1-6\xi\Phi_\xi^2}$, with these two terms counteracting each
other.  Moreover, it can be seen that $g_{tt}>0$ for ${\bar r} \geq
\eta/2 $, that is, we do not find any horizon by going inward from the
asymptotic region. In fact, we can extend these geometries to values
${\bar r} < \eta/2 $ but we will leave for the next section to
describe the traversable wormhole nature of these solutions.

The solutions with $\chi\in(0, \pi/3)$ also deserve some additional
attention. They are wormhole-like shaped, that is, they have two
asymptotic regions (at ${\bar r}=\infty $ and ${\bar r}=\eta/2$)
joined by a throat. However, analyzing the $tt$ component of the Ricci
tensor ($R_{\hat t\hat t}$ in an orthonormal coordinate basis) one can
easily realize that it diverges when the scalar field reaches the
value $-1/\sqrt{6\xi}$, that is, at ${\bar r}=\eta/2$.  Therefore, the
region ${\bar r} \rightarrow \eta/2$ is not a proper ``asymptotic''
region. (See the discussion on this point in~\cite{barcelo}).
Although in these geometries there exist diverging-lens effects (they
have a throat), they are not genuine traversable wormholes.

The solutions with $F_{-}$ in equation (\ref{E:src}) merit a
discussion similar to that made for the plus sign. The real function
$F_{-}H$ is only defined for absolute values of the scalar field
greater than $1/\sqrt{6\xi}$, and of course less than
$1/\sqrt{6\xi(1-6\xi)}$. (See figure 3.)  When the scalar field reaches
a value slightly lower than $-1/\sqrt{6\xi}$ the radial coordinate
${\bar r}$ approaches $\eta/2$, with $F_{-}H$ going to zero. At this
coordinate point we can perform the same analysis as before,
concerning the different behaviour as a function of $\chi$ of the
Schwarzschild radial coordinate ${\cal R}_{-}$.  Obviously, we find
the same results. The case in which we are most interested is that of
$\chi=\pi/3$. In the next section we will see how by extending the
geometry beyond $\eta/2$ we finally get a perfectly well-defined
traversable wormhole.  However, for this to happen it is necessary
that by decreasing the value of the scalar field from the critical
value $-1/\sqrt{6\xi}$, one must reach an asymptotic region before
arriving to the lowest possible value of $-1/\sqrt{6\xi(1-6\xi)}$. That
can only be guaranteed if the condition
\begin{equation}
\Phi_{-}>{\rm exp}\left(-\sqrt{{1-6\xi \over 6\xi}} \; {\pi\over2}\right)
\label{E:condition-} 
\end{equation} 
is fulfilled. This condition is obtained by requiring that $F_{-}H$
have a value greater than one for $\Phi=-1/\sqrt{6\xi(1-6\xi)}$.  In
the example of figure 3 this is not satisfied. This constraint implies
super-Planckian values of the scalar field in the wormhole throat, and
is a cause for some mild concern --- we shall return to this point
shortly.

\subsection{Solutions with $\xi=1/6$}

The coupling constant $\xi=1/6$ corresponds to a conformal coupling
prescription.  We have already found all the classical solutions for this
system in~\cite{barcelo}, highlighting the many interesting
characteristics possessed by the conformal coupling.  Here, we present
these solutions as a particular case of the curvature coupling.  For this
particular case the function $H$ is unity and the functions $F_{\pm}$
become
\begin{equation}
F_{\pm}(\Phi_{[\xi=1/6]})
=
\Phi_\pm 
\sqrt{\pm { 1+\Phi_{[\xi=1/6]} \over 1-\Phi_{[\xi=1/6]} }}.
\end{equation}
The expression (\ref{E:fh}) can be inverted\footnote{
Tip: remember that ${1 \over 2}\ln {1+x \over 1-x}=\tanh^{-1}(x)$.}
yielding the two solutions (\ref{E:physicalsol1}) and
(\ref{E:physicalsol2}), provided we identify $\Phi^0_{-}=\ln \Phi_{-}$
and $\Phi^0_{+}=\ln \Phi_{+}$.

The analysis done for the previous $0<\xi<1/6$ case extends directly
to $\xi=1/6$. The special geometry corresponding to $\chi=\pi/3$
will be described more fully in the next section in combination with
the equivalents for the range $0<\xi<1/6$.

\subsection{Solutions with $\xi>1/6$}

Once more it is necessary to take into account the functions $F_{+}$
and $F_{-}$ separately.  The form of $F_{+}H$ tells us that the scalar
field reaches some finite value in an asymptotic region, at the point
with $F_{+}H=1$.  (See figure 4.)  Then, by going inward from this
asymptotically flat region we approach the ${\bar r}= \eta/ 2$
section, for a value of the scalar field equal to $-1/\sqrt{6\xi}$.
Perturbatively analyzing the form of $F_{+}H$ around
$\Phi_\xi=-1/\sqrt{6\xi}$ we can see that ${\cal R}_{+}$ has a
different behaviour depending on the value of $\chi$.  One again
obtains the same qualitative results as for the previous $0<\xi<1/6$
case. For $0<\chi<\pi/3$ the geometry blows up in a singular way, for
$\pi/3<\chi<\pi$ the geometry shrinks to a naked singularity, and in
the special $\chi=\pi/3$ case the geometry can be extended to values
${\bar r} <\eta/ 2$.  Once more, this particular case will be
described in the next section.

If we now consider the function $F_{-}H$, we can see that it is only
defined for $|\Phi_\xi|>1/\sqrt{6\xi}$. Here, for a negative value of
the field lower than $-1/\sqrt{6\xi}$ we have an asymptotically flat
region, ${\bar r} \rightarrow \infty$ ($F_{-}H \rightarrow 1$), but
now this happens independently of the value of $\Phi_{-}$. (See figure
5.)  Increasing the value of the scalar field up to $-1/\sqrt{6\xi}$
we arrive at the inner ${\bar r} =\eta/ 2$ region.  At this coordinate
point we find once more what we now see are the three possible
standard behaviours for the geometry depending on $\chi$.

\section{Non-minimal scalars and traversable wormholes}
\label{S:wormhole}
In this section we shall describe in more detail the solutions
with $\chi=\pi/3$ and $\xi>0$.  In all these cases the expression
under the square root symbol of the functions $F_{+}$ or $F_{-}$
changes its sign when $\Phi_\xi=-1/\sqrt{6\xi}$ (${\bar r} =\eta/ 2$).
It seems that we cannot define the scalar field beyond this point
because the apparently complex values that $F$ would generate.  However,
we have to realize that for $\chi=\pi/3$ the expression
(\ref{E:implicitrbar}) can be written as
\begin{equation}
\left[ {1- {\eta \over 2 {\bar r}} \over 
1+ {\eta \over 2 {\bar r}}} \right]=
[F(\Phi_\xi)H(\Phi_\xi)]^2.
\label{E:implicitrbarp3}
\end{equation}
which we shall see allows us to avoid the problem and permits the
extension.  By using isotropic coordinates and particularizing to the
value $\chi=\pi/3$ one can realize that there is a convenient way in
which we can write the metric:
\begin{equation}
ds_\xi^2=\pm{(F_{\pm}H)^2 \over 1-6\xi \Phi_\xi^2 }
\left[ -dt^2+\left(1+{\eta \over 2 {\bar r}} \right)^4 [d{\bar r}^2+
{\bar r}^2 (d\theta^2+\sin^2 \theta \; d\Phi^2)] \right].
\label{E:wh}
\end{equation}
Let us see what happens in the $0<\xi<1/6$ case when we choose the
function $F_{+}$.  As we explained in the previous section, the
implicit relation (\ref{E:implicitrbarp3}), and the form of the
Schwarzschild radial coordinate (\ref{E:src}), tell us that we have a
geometry that is perfectly regular from ${\bar r} \rightarrow \infty$
to ${\bar r}=\eta/2$, at which point the scalar field acquires the
value $-1/\sqrt{6\xi}$. The function $(F_{+}H)^2$ has a zero for
$\Phi_\xi=-1/\sqrt{6\xi}$ which counteracts the zero in the
denominator of the prefactor in equation (\ref{E:wh}). Therefore,
despite naive appearances, this prefactor acquires a finite positive
value on $\Phi_\xi=-1/\sqrt{6\xi}$.  Beyond that point, that is for
${\bar r}<\eta/2$ and $\Phi<-1/\sqrt{6\xi}$, we can see that this
prefactor continues to be finite and positive [$(F_{+}H)^2$ changes
its sign in the same way as $(1-6\xi \Phi_\xi^2)$] even at the lowest
possible value for the scalar field $-1/\sqrt{6\xi(1-6\xi)}$.  Now, if
the radial coordinate ${\bar r}$ reaches the value zero before the
scalar field reaches its lowest possible value
$-1/\sqrt{6\xi(1-6\xi)}$, that is, if the condition
\begin{equation}
\Phi_{+}>{\rm exp}\left(- \sqrt{{1-6\xi \over 6\xi}} \;{\pi\over2} \right)
\label{E:condition+}
\end{equation}
is fulfilled, expression (\ref{E:wh}) tells us that for ${\bar r}=0$
there is another asymptotically flat region.  This spacetime is now a
perfectly well-defined traversable wormhole geometry.

If the condition (\ref{E:condition+}) is {\em not} satisfied, then the
scalar field reaches the critical value $-1/\sqrt{6\xi(1-6\xi)}$
before the function $(F_{+}H)^2$ reaches the value $-1$ for which the
other asymptotic region shows up.  In these solutions, we can extend
the geometry and the scalar field regularly up to a spherical section
at which the scalar field approaches the value
$-1/\sqrt{6\xi(1-6\xi)}$.  At this value we can see, by
differentiating (\ref{E:implicitrbar}) with respect ${\bar r}$, that
the derivative of the scalar field becomes infinite. This, and the
form of the scalar curvature for these systems,
\begin{equation}
R_\xi={ 6(\nabla\Phi_\xi)^2  \; (1-6\xi)\over 1-6\xi(1-6\xi)\; \Phi_\xi^2 },
\label{E:r}
\end{equation}
tells us that there is a curvature singularity in these solutions.
(This will be a naked singularity hiding behind a wormhole throat,
but the lack of a second asymptotic region implies these are not true
traversable wormhole solutions.)
 
We can now realize that by choosing the $F_{-}$ branch and extending
upwards from scalar field values greater than $-1/\sqrt{6\xi(1-6\xi)}$
towards $\Phi_\xi>-1/\sqrt{6\xi}$, we obtain the same solutions as for
$F_{+}$, but with interchanged asymptotic regions. The prefactor in
equation (\ref{E:wh}) is $(F_{-}H)^2/(6\xi\;\Phi_\xi^2-1)$ in this
case.  The condition (\ref{E:condition-}) now plays the same role here
as that of (\ref{E:condition+}) previously.

The traversable wormhole solutions that we found in~\cite{barcelo} for
the conformal case can be rediscovered here as the limiting case of
those with $0<\xi<1/6$. The conditions (\ref{E:condition-}) and
(\ref{E:condition+}) for the existence of wormholes, become
$\Phi_{-}>1$ and $\Phi_{+}>1$ which yield easily the conditions
$\Phi^0_{-}=\ln \Phi_{-}>0$ and $\Phi^0_{+}=\ln \Phi_{+}>0$ for the
conformal wormholes.

In the case $\xi>1/6$, and using the function $F_{+}$, we had
solutions with an asymptotic region in which the scalar field acquired
a value greater than $-1/\sqrt{6\xi}$.  From this asymptotic region
one can move in the direction in which the scalar field decreases. In
this way, first one crosses the non-singular section with
$\Phi_\xi=-1/\sqrt{6\xi}$.  Later on, one crosses a spherical section
with a minimum diameter (a throat), and from that point on the size of
the spherical sections begin to grow to reach another asymptotic
region associated with some asymptotic value for the scalar field
which is of course lower than $-1/\sqrt{6\xi}$. Here, one does not
have to impose any restriction on $\Phi_{+}$ in order to obtain a
traversable wormhole configuration.

By choosing the function $F_{-}$ we find the same solutions,
but coordinatized in a reversed way: The asymptotic region with
a greater scalar field value is here that with ${\bar r}=0$.  

In view of the plethora of traversable wormhole solutions we have
found, an important observation is in order. In all these solutions
the scalar field has to reach absolute values above $\sim
m_p/\sqrt{\xi}$, where $m_p$ is the Planck mass. That is, either the
scalar field acquires trans-Planckian values or the curvature coupling
constant $\xi$ must become disturbingly large\footnote{
This might suggest that ultimately a proper quantum treatement of 
these wormholes would be desirable.}.
Moreover, imagining
that we include in the system some additional matter field with
an action
\begin{equation}
\S^m=\int \sqrt{-g_{\xi}} \; f(\Phi_\xi) \; \L^m,
\end{equation}
its contribution to the effective energy-momentum tensor would be
\begin{equation}
[T^m_{\eff}]_{\mu\nu}
=
{f(\Phi_\xi) \over 1-6\xi \;\Phi^2_\xi} \; T^m_{\mu\nu}.
\end{equation}
This behaves {\em as if} this matter interacts
through an ``effective Newton constant''
\begin{equation}
\tilde{G}_{\eff}=G_N \; {f(\Phi_\xi) \over 1-6\xi \; \Phi^2_\xi} = 
{1\over8\pi} \; {f(\phi_\xi) \over \kappa-\xi \; \phi^2_\xi}.
\label{effective}
\end{equation}
(This is not yet quite the physical Newton constant even in the 
standard case $f(\Phi_\xi)=1$; see section \ref{S:discussion} below). 
Then, unless $f(\Phi_\xi)$ is
specifically chosen to counteract the $1-6\xi \; \Phi^2_\xi$ factor,
this effective Newton constant would change its sign form one
asymptotic region to the other, producing weird effects on ordinary
matter\footnote{
It might be also a source of problems for the stability of the 
solutions found, but we have not addressed this problem here.}.
One can try to ``build'' symmetric wormholes beginning from
these asymmetric wormholes~\cite{barcelo} and performing ``thin-shell
surgery''~\cite{examples}.  In this way, one could potentially
restrict the peculiar effects on matter to a thin region around the
wormhole throat.

\section{Summary and discussion}
\label{S:discussion}

We have seen that a non-minimally coupled scalar field can violate all
the energy conditions at a classical level. The violation, in
principle, of the ANEC inspired us to look for the possible existence
of traversable wormhole geometries supported by these non-minimal
scalar fields.  We have obtained the classical solutions for gravity
plus a massless arbitrarily coupled scalar field in the restricted
class of spherically symmetric and static configurations.  Let
separate the different cases
\begin{enumerate}
\item[{\it \bf i)}] $\xi<0$:

We find naked singularity geometries and, in the case of a constant
value for the scalar field, we recover the Schwarzschild solution. 

\item[{\it \bf ii)}] $\xi=0$:

In this case we have the usual class of minimally coupled solutions
with its naked singularities. Again, we recover the Schwarzschild
solution for a constant field.

\item[{\it \bf iii)}] $0<\xi<1/6$:

We find assorted naked singularities. In some of them the geometry
shrinks to zero, in others the singularity is placed in an asymptotic
region, and yet others there is a scalar curvature singularity at a
finite size spherical section.

For a constant value of the field we recover the Schwarzschild
solution, but if this value is exactly $\pm 1/\sqrt{6\xi}$ the
geometry can be arbitrarily chosen from among the solutions for the
conformal $1/6$ case.

Apart from these solutions we find a two-parameter family of perfectly
well-defined traversable wormholes geometries. Given an asymptotic
value for the scalar field, with absolute value lower than
$1/\sqrt{6\xi(1-6\xi)}$, and a scalar charge, the asymptotic mass for
a traversable wormhole is fixed.

\item[{\it \bf iv)}] $\xi=1/6$:
 
The case of conformal coupling leads to the same type of naked
singularities that were seen before, but the scalar curvature
singularities are now absent.

The different geometries can appear either in combination with a
suitable spatial dependent scalar field, or with a constant field
$\Phi=\pm 1$.

Here, we have also a two-parameter family of traversable wormholes
with the only difference that now the asymptotic value of the scalar
field can have arbitrary values.  In fact, when one asymptotic region
has an infinite value for the scalar field it means that it is not
asymptotically flat, degenerating to a cornucopia~\cite{barcelo}.

\item[{\it \bf v)}] $\xi>1/6$:

Once more we find assorted naked singularities of the same type
as in the conformal coupling.

For a special constant value of the field, $\pm 1/\sqrt{6\xi}$, the
geometry can again be arbitrarily selected from among the solutions
for the conformal $1/6$ case.  For a different constant value we only
recover the Schwarzschild solution.

Also, there are specific traversable wormhole solutions in which an
infinite asymptotic value for the scalar field cannot be reached.
\end{enumerate}
Since the potential existence of traversable wormholes is a perhaps
somewhat disturbing possibility, we feel it a good idea to see where
the potential pitfalls might be --- mathematically, we have exhibited
exact classical traversable wormhole solutions to the Einstein
equations, and now wish to investigate the extent to which they should
physically be trusted.

To start with, scalar fields play a somewhat ambiguous role in modern
theoretical physics: on the one hand they provide great toy models,
and are from a theoretician's perspective almost inevitable components
of any reasonable model of empirical reality; on the other hand the
direct experimental/observational evidence is spotty.

The only scalar fields for which we have really direct ``hands-on''
experimental evidence are the scalar mesons (pions $\pi$; kaons $K$;
and their ``charm'', ``truth'', and ``beauty'' relatives, plus a whole
slew of resonances such as the $\eta$, $f_0$, $\eta'$,
$a_0$,\dots)~\cite{PDG}.  Not a single one of these particles are
fundamental, they are all quark-antiquark bound states, and while the
description in terms of scalar fields is useful when these systems are
probed at low momenta (as measured in their rest frame) we should
certainly not continue to use the scalar field description once the
system is probed with momenta greater than
$\hbar/({\mathrm{bound~state~radius}})$. In terms of the scalar field
itself, this means you should not trust the scalar field description
if gradients become large, if
\begin{equation}
||\nabla \phi|| > {||\phi||\over {\mathrm{bound~state~radius}} }.
\end{equation}
Similarly you should not trust the scalar field description if the
energy density in the scalar field exceeds the critical density for
the quark-hadron phase transition. (Note that if the scalar mesons
were strict Goldstone bosons [exactly massless] rather than
pseudo--Goldstone bosons, then they could achieve arbitrarily large
values of the field variable with zero energy cost.) Thus scalar
mesons are a mixed bag: they definitely exist, and we know quite a bit
about their properties, but there are stringent limitations on how far
we should trust the scalar field description.

The next candidate scalar field that is closest to experimental
verification is the Higgs particle responsible for electroweak
symmetry breaking.  While in the standard model the Higgs is
fundamental, and while almost everyone is firmly convinced that some
Higgs-like scalar field exits, there is a possibility that the
physical Higgs (like the scalar mesons) might itself be a bound state
of some deeper level of elementary particles (\eg, technicolor and its
variants). Despite the tremendous successes of the standard model of
particle physics we do not (currently) have direct proof of the
existence of a fundamental Higgs scalar field. 

Accepting for now the existence of a fundamental Higgs scalar, what is
its curvature coupling? The parameter $\xi$ is completely
unconstrained in the flat-space standard model. If we choose for
technical reasons to adopt the ``new improved stress energy tensor''
for the Higgs scalar in flat spacetime then one is naturally led to
conformal coupling in curved spacetime~\cite{barcelo}. Conformal
coupling seems to be the ``most natural'' choice for the Higgs
curvature coupling, and we have seen in this note that both conformal
coupling and the entire open half-line $\xi\in(0,\infty)$ surrounding
conformal coupling lead to traversable wormhole geometries.
Unfortunately adding a Higgs mass results in analytically intractable
equations.

A third candidate scalar field of great phenomenological interest is
the axion: it is extremely difficult to see how one could make strong
interaction physics compatible with the observed lack of strong CP
violation, without something like an axion to solve the so-called
``strong CP problem''. Still, the axion has not yet been directly
observed experimentally.

A fourth candidate scalar field of phenomenological interest
specifically within the astrophysics/cosmology community is the
so-called ``inflaton''. This scalar field is used as a mechanism for
driving the anomalously fast expansion of the universe during the
inflationary era. While observationally it is a relatively secure bet
that something like cosmological inflation (in the sense of
anomalously fast cosmological expansion) actually took place, and
while scalar fields of some type are presently viewed as the most
reasonable way of driving inflation, we must again admit that direct
observational verification of the existence of the inflaton field (and
its variants, such as quintessence) is far from being accomplished.
Note that in many forms of inflation trans--Planckian values of the
scalar field are generic and widely accepted (though not universally
accepted) as part of the inflationary paradigm.

A fifth candidate scalar field of phenomenological interest
specifically within the general relativity community is the so-called
``Brans--Dicke scalar''. This is perhaps the simplest extension to
Einstein gravity that is not ruled out by experiment. (It is certainly
greatly constrained by observation and experiment, and there is no
positive experimental data guaranteeing its existence, but it is not
ruled out.) The relativity community views the Brans--Dicke scalar
mainly as an excellent testing ground for alternative ideas and as a
useful way of parameterizing possible deviations from Einstein
gravity. (And experimentally and observationally, Einstein gravity
still wins.)

In this regard it is important to emphasize that the type of scalar
fields we have been discussing in the present paper are completely
compatible with current experimental limits on Brans--Dicke scalars,
or more generally, generic scalar-tensor theories~\cite{will,weinberg}.
To take the observational limits and utilize
them in our formalism you need to use the translations
\begin{equation}
\Phi_{\mathrm{Brans-Dicke}} = \kappa - \xi \;\phi_\xi^2.
\end{equation}
\begin{equation}
\omega(\Phi_{\mathrm{Brans-Dicke}}) 
= 
{
\Phi_{\mathrm{Brans-Dicke}} 
\over 
\left( \d \Phi_{\mathrm{Brans-Dicke}}/ \d \phi_\xi \right)^2 
}
={\kappa - \xi \; \phi_\xi^2\over4\; \xi^2\; \phi_\xi^2}.
\end{equation}
Then, the physical Newton constant, which takes its meaning
within the perturbative PPN expansion around flat Minkowski space, 
can be written as
\begin{equation}
G_{\mathrm{physical}} 
= 
{1\over\Phi_{\mathrm{Brans-Dicke}}} \; 
\left( { 4 + 2\omega \over 3 + 2\omega } \right) \bigg|_{\rm asymp}
=
{1\over8\pi} \; 
{1 \over \kappa-\xi \; \phi^2_\xi} \;
\left( { 4 + 2\omega \over 3 + 2\omega } \right) \bigg|_{\rm asymp}.
\end{equation}
(Here, we have set the function $f(\phi_\xi)$ in (\ref{effective}) 
equal to one. However it must be pointed out that if 
$f(\phi_\xi)\neq 1$ then the current theories are more general even 
than the standard scalar-tensor models). 
In the wormhole solutions that we have found, the scalar field $\phi_\xi$ 
reaches 
a trans-Planckian value in one of the asymptotic regions, making 
the physical Newton constant negative. This means that, contrary to
what is commonly done in scalar-tensor theories~\cite{santiago},
the translated Brans-Dicke field should not be restricted to only 
positive values. In order to cover our wormhole solutions, negative 
values of the
Brans-Dicke field are required.

Current solar system measurements imply
$||\omega(\Phi_{\mathrm{Brans-Dicke}})|| > 3000$~\cite{vlbi}. More
precisely, VLBI techniques allow us to use the deflection of light by
the Sun to place very strong constraints on the PPN parameter
$\gamma$, with~\cite{vlbi}
\begin{equation}
\gamma = 1-(0.6\pm3.1) \times 10^{-4}.
\end{equation}
The standard result that for Brans--Dicke
theories~\cite{will,weinberg}
\begin{equation}
\gamma = {\omega +1 \over \omega+2}
\end{equation}
now provides the limit on $||\omega||$ quoted above. For instance, 
this constraint is very easily satisfied if the scalar field $\phi_\xi$
is a small fraction of the naive Planck scale ($\sqrt{\kappa}$) in 
the solar neighborhood. Thus, the scalar theories we investigate
in this paper are perfectly compatible with known physics.

Finally, the membrane-inspired field theories (low-energy limits of
what used to be called string theory) are literally infested with
scalar fields. In membrane theories it is impossible to avoid scalar
fields, with the most ubiquitous being the so-called ``dilaton''.
However, the dilaton field is far from unique, in general there is a
large class of so-called ``moduli'' fields, which are scalar fields
corresponding to the directions in which the background spacetime
geometry is particularly ``soft'' and easily deformed. So if membrane
theory really is the fundamental theory of quantum gravity, then the
existence of fundamental scalar fields is automatic, with the field
theory description of these fundamental scalars being valid at least
up to the Planck scale, and possibly higher.

(For good measure, by making a conformal transformation of the
spacetime geometry it is typically possible to put membrane-inspired
scalar fields into a framework which closely parallels that of the
generalized Brans--Dicke fields. Thus there is a potential for much
cross-pollination between Brans--Dicke inspired variants of general
relativity and membrane-inspired field theories.)

So overall, we have excellent theoretical reasons to expect that
scalar field theories are an integral part of reality, but the direct
experimental/observational verification of the existence of
fundamental fields is still an open question. Nevertheless, we think
it fair to say that there are excellent reasons for taking scalar
fields seriously, and excellent reasons for thinking that the
gravitational properties of scalar fields are of interest
cosmologically, astrophysically, and for providing fundamental probes
of general relativity. The fact that scalar fields then lead to such
widespread violations of the energy
conditions~\cite{flanagan,barcelo,visser3,bekenstein-bounce}, with
potentially far-reaching consequences like a ``universal bounce''
(instead of a big-bang
singularity)~\cite{bekenstein-bounce,bounce1,bounce2}, the traversable
wormholes of this paper (see also~\cite{barcelo,visser3}),
and possibly even weirder physics (see for example~\cite{visser1}),
leads to a rather sobering assessment of the marked limitations of our
current understanding.

\section*{Acknowledgments}
The research of CB was supported by the Spanish Ministry of Education
and Culture (MEC). MV was supported by the US Department of Energy.
MV wishes to thank Jacob Bekenstein for his interest and comments.

 

\begin{figure}[htb]
\vbox{ 
\hfil
\scalebox{0.4}{\includegraphics{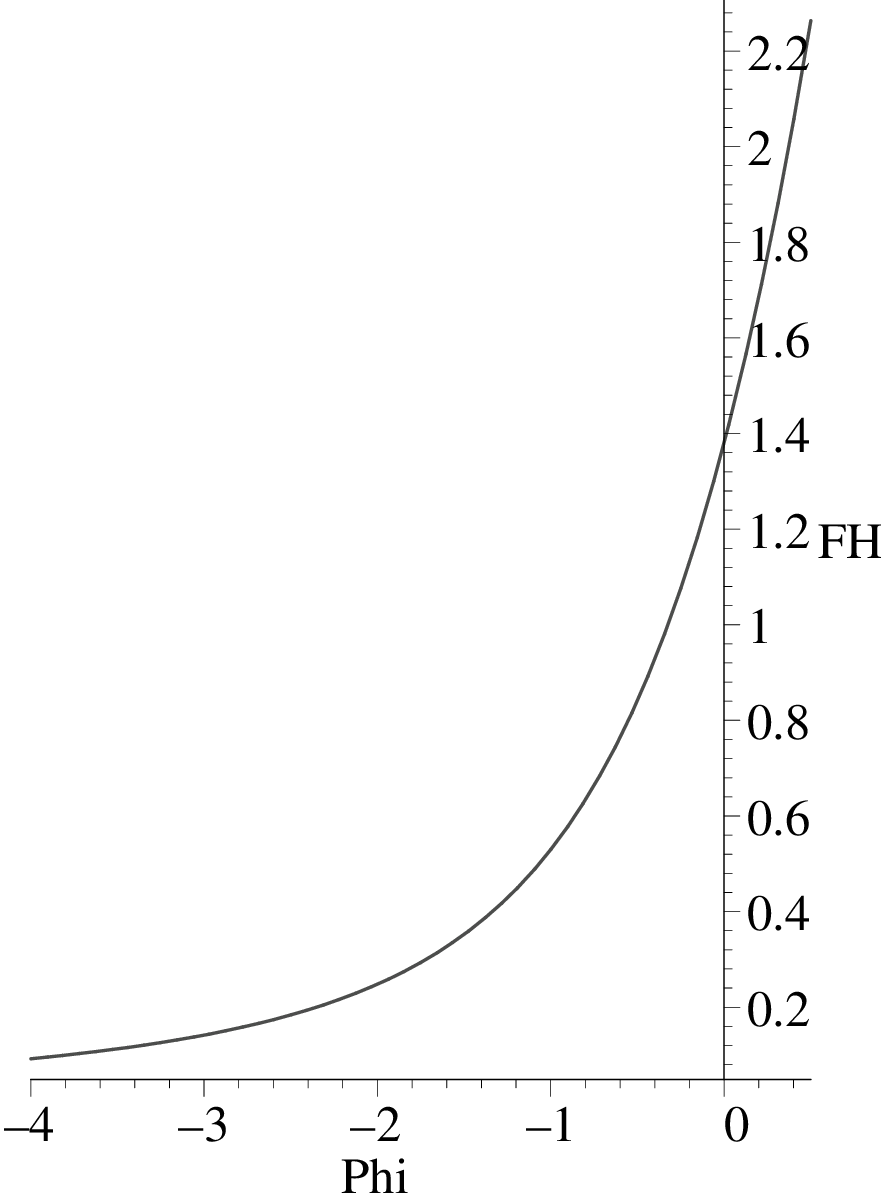}}
\hfil 
}
\bigskip
\caption{\label{F:fig1} $F_{+}H$, $\xi=-0.2$, $\chi=\pi/6$. 
There are no traversable wormholes for $\xi<0$, and the solutions
generically contain naked singularities. }
\end{figure}

\begin{figure}[htb]
\vbox{ 
\hfil
\scalebox{0.4}{\includegraphics{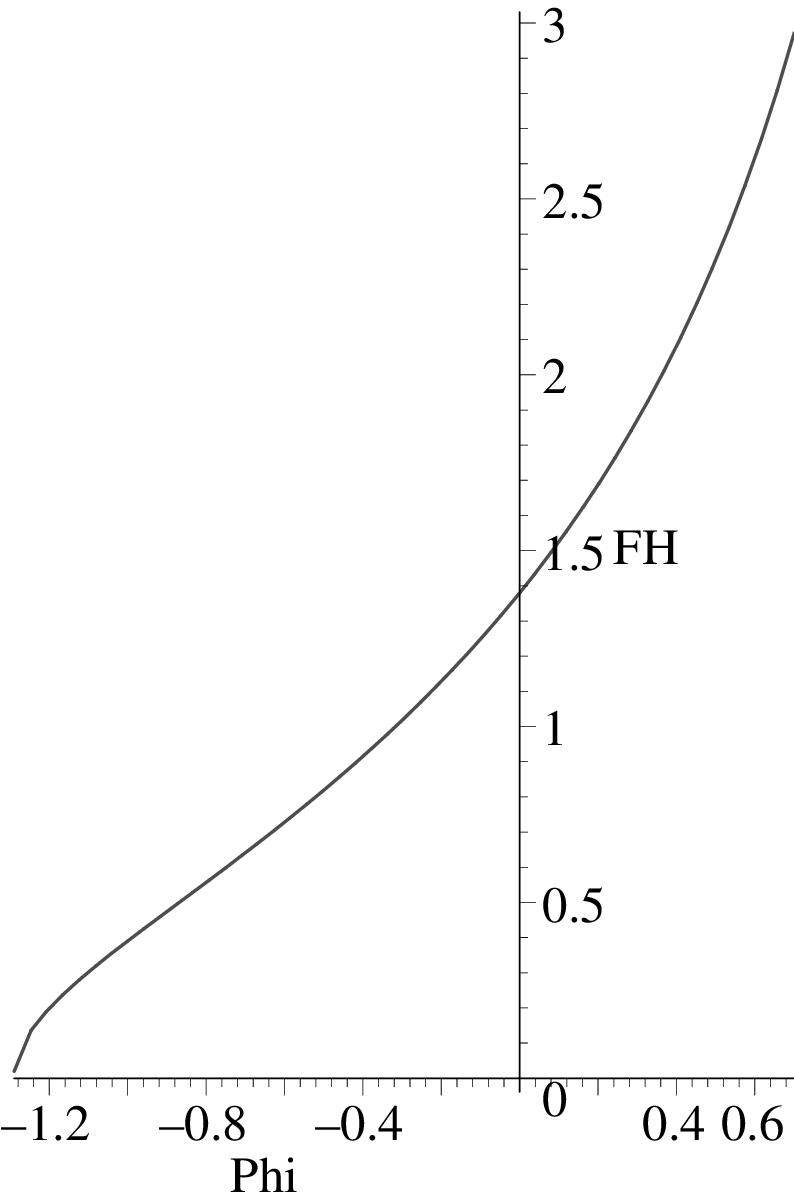}}
\hfil 
}
\bigskip
\caption{\label{F:fig2} $F_{+}H$, $\xi=0.1$, $\chi=\pi/6$.
For $\xi\in(0,1/6)$ both the $F_+$ and $F_-$ branches need to be considered. }
\end{figure}

\begin{figure}[htb]
\vbox{ 
\hfil
\scalebox{0.4}{\includegraphics{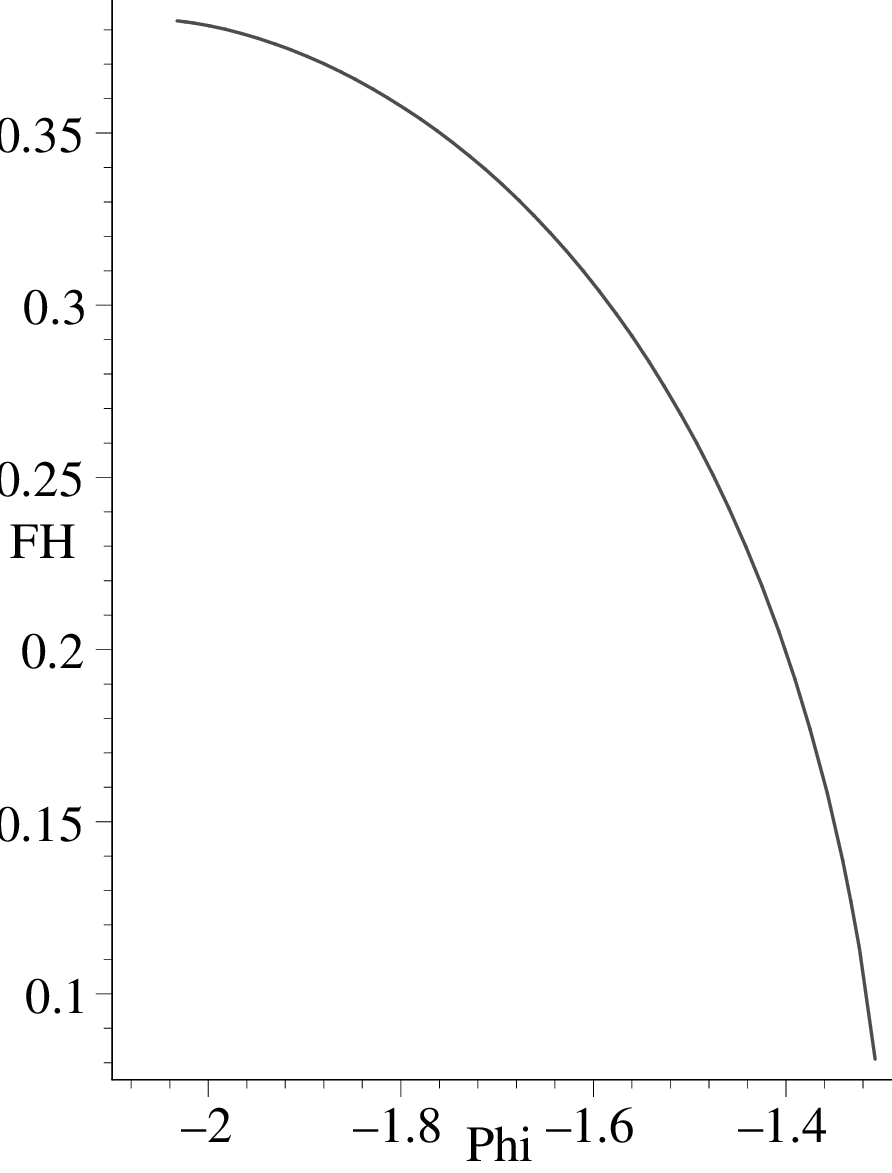}}
\hfil 
}
\bigskip
\caption{\label{F:fig3} $F_{-}H$, $\xi=0.1$, $\chi=\pi/6$.
For $\xi\in(0,1/6)$ both the $F_+$ and $F_-$ branches need to be considered.  }
\end{figure}

\begin{figure}[htb]
\vbox{ 
\hfil
\scalebox{0.4}{\includegraphics{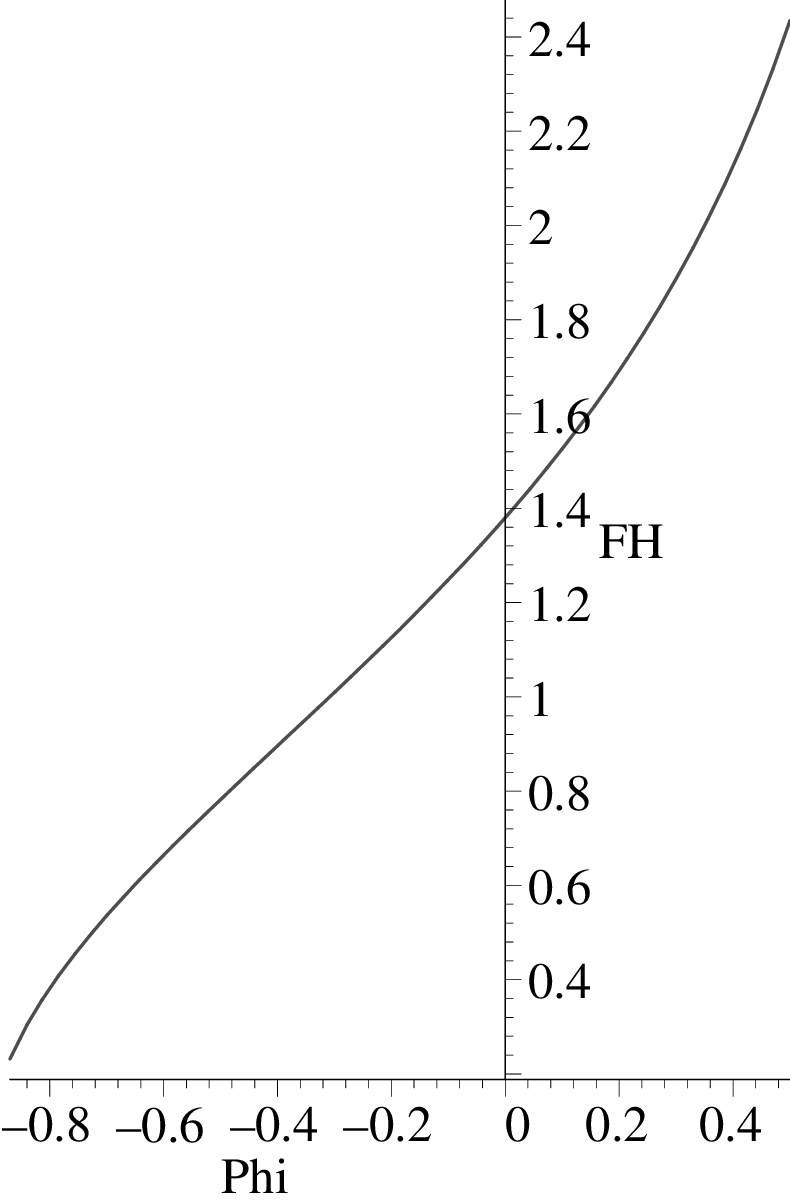}}
\hfil 
}
\bigskip
\caption{\label{F:fig4} $F_{+}H$, $\xi=0.2$, $\chi=\pi/6$.
For $\xi\in(0,1/6)$ both the $F_+$ and $F_-$ branches need to be considered. }
\end{figure}

\begin{figure}[htb]
\vbox{ 
\hfil
\scalebox{0.4}{\includegraphics{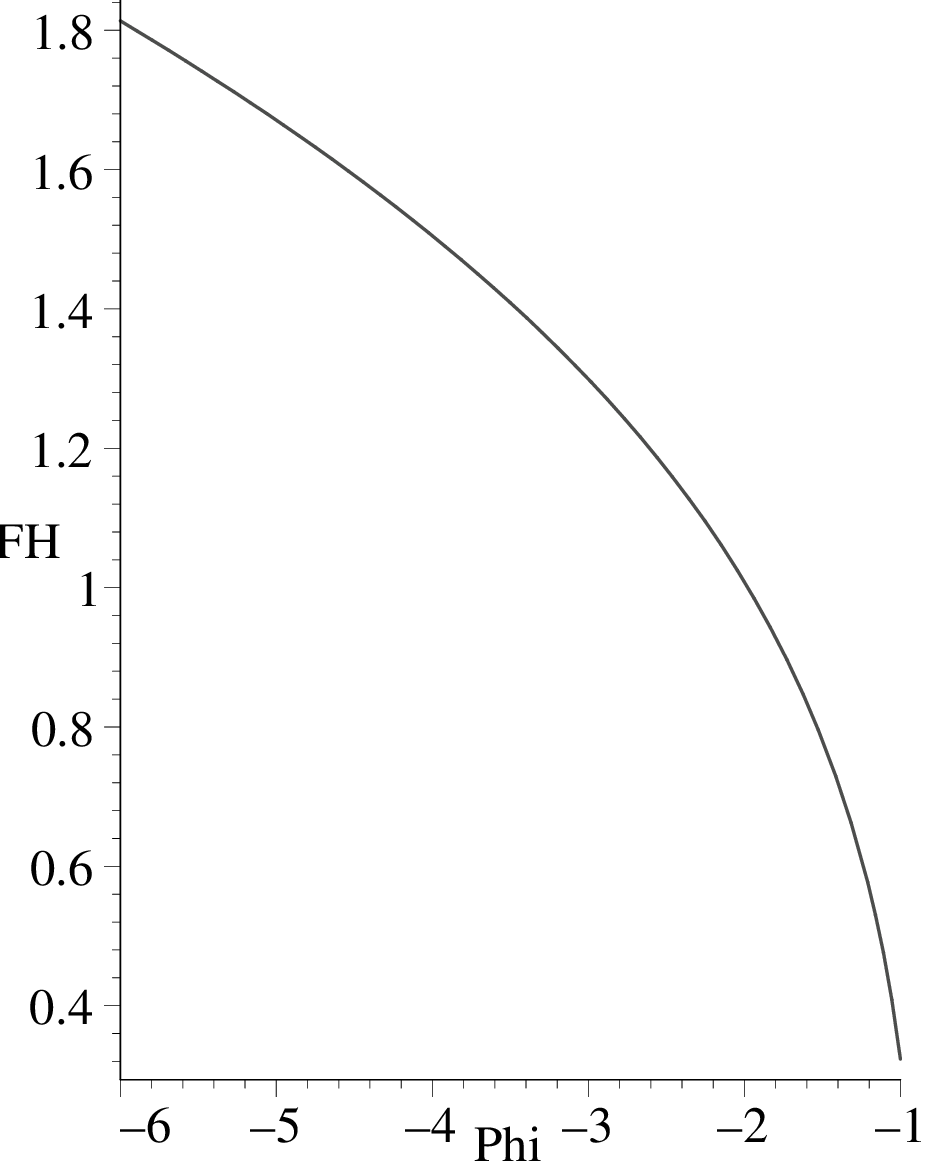}}
\hfil 
}
\bigskip
\caption{\label{F:fig5} $F_{-}H$, $\xi=0.2$, $\chi=\pi/6$.
For $\xi\in(0,1/6)$ both the $F_+$ and $F_-$ branches need to be considered. }
\end{figure}

\end{document}